# Quantum Crystallography:
# *N*-Representability[†‡]

## Chérif F. Matta [1,2,3,4], Lulu Huang [5], and Lou Massa [5,6,*]

[1] *Department of Chemistry and Physics, Mount Saint Vincent University, Halifax, Nova Scotia, Canada B3M 2J6*
[2] *Dalhousie University, Halifax, Nova Scotia, Canada B3H 4J3*
[3] *Saint Mary's University, Halifax, Nova Scotia, Canada B3H 3C3*
[4] *Département de chimie, Université Laval, Québec, Québec, Canada G1V 0A6*
[5] *Department of Chemistry, Hunter College, City University of New York, NY 10065, USA*
[6] *Departments of Chemistry and Physics, Graduate Center, City University of New York, NY 10016, USA*

**\*** Correspondence: lmassa@hunter.cuny.edu

[†] This paper is to honor the memory of Linus Pauling (1901-1994) on the occasion of his 120th birthday (1901).
[‡] The results presented in this paper are patterned upon a presentation by L.M. entitled "Quantum Crystallography: *N*-Representability Big and Small" during the *Premier Quantum Crystallography Online Meeting* (QCrOM2020) at CentraleSupélec, Paris, 26-29 August 2020.

## GRAPHICAL ABSTRACT

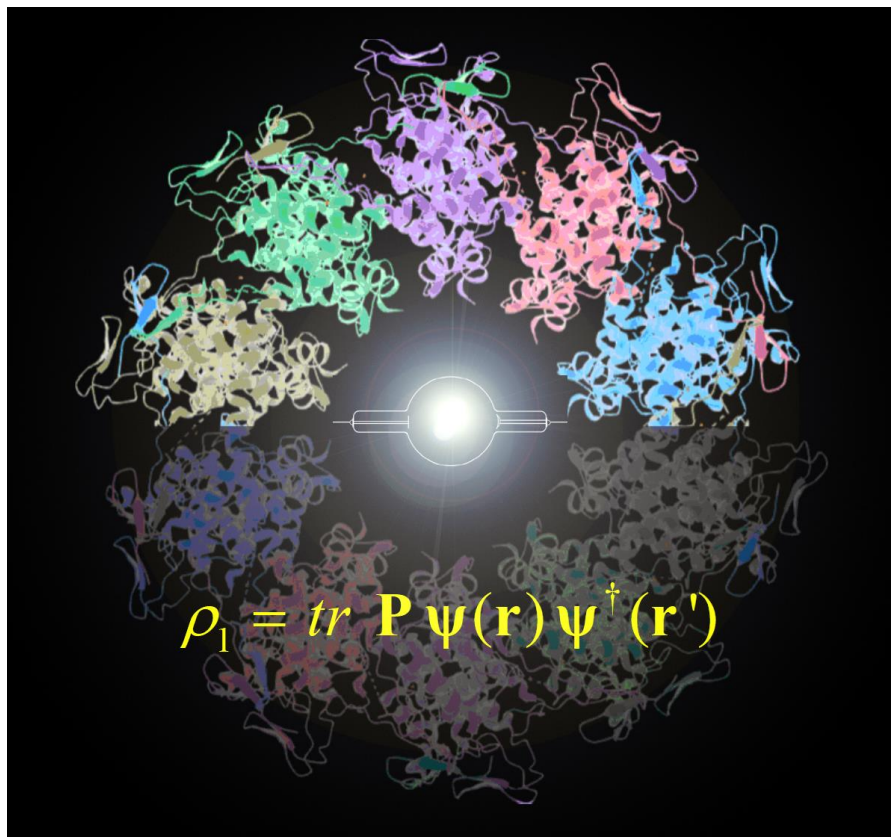

$\rho_1 = tr\, \mathbf{P}\, \psi(\mathbf{r})\psi^\dagger(\mathbf{r}')$

**Abstract:** Linus Pauling contributions span structural biology, chemistry in its broadest definition, quantum mechanical theory, valence bond theory, and even nuclear physics. A principal tool developed and used by Pauling is X-ray (and electron) diffraction. One possible extension of Pauling's oeuvre could be the "marriage" of crystallography and quantum mechanics. Such an effort dates back to the sixties and has now flourished into an entire subfield termed "*quantum crystallography*". Quantum crystallography could be achieved through the application of Clinton equations to yield *N*-representable density matrices consistent with experimental data. The implementation of the Clinton equations is qualitatively different for small and for large systems. For a small system, quantum mechanics is *extracted* from X-ray data while for a large system, the quantum mechanics is *injected* into the system. In both cases, *N*-representability is imposed by the use of the Clinton equations.

1.      Introduction

Linus Pauling, one of the greatest chemists, was in some sense the embodiment of quantum crystallography broadly defined. Pauling advanced quantum mechanics *per se*[1] and its applications as diverse as in the development of valence bond (VB) theory[2,3] and in elucidating the "nature of the chemical bond",[4] all the way through studies of atomic nuclei (see his thematically-sorted list of publications appended to Ref. 5). Pauling's invention of directed hybrid orbitals was meant to mimic the directionality of chemical bonds discovered by crystallography. Pauling has also been a towering crystallographer, using both electron diffraction and X-ray techniques, whose work has provided us, for example, with much of what we know today regarding protein higher-order structures (secondary to tertiary).

Quantum crystallography, in its modern definition,[6-18] includes the effort of recovering electronic structure from diffraction data. This would enable the study of the "nature of the chemical bond" directly from the result of experiment when constrained by theory. This, we surmise, would have been a delightful development for Pauling.

One important feature of a quantum mechanically derived electron density (matrices) is that it satisfies *N*-representability. In other words, such density (matrices) must be mappable into the function space of antisymmetric wavefunctions – restricting considerably those which are legitimate. Joining both ends of the Pauling oeuvre, and that is, an attempt to provide electronic description of a large biopolymer of which the structure is known, one has to resort to some form of a fragmentation scheme. Fragmentation is necessary since one quickly hits the computational bottleneck that is well-known to plague molecular quantum mechanical calculations. A question arises regarding the *N*-representability of the "reassembled" electron density (matrices) reconstructed from fragments. Constraining the reconstructed density (matrices) to *N*-representability guaranties its/their proper quantum mechanical description.

This paper explores the possibility of obtaining *N*-representable electron density (and reduced density matrices) within the one-determinant approximation of arbitrarily large biopolymers. This is achieved within the framework of the kernel energy method (KEM) of fragmentation.[14] As a prelude, and to make this paper self-contained, we start by reviewing some basic concepts.

## 2.     N-Representability

An electron density matrix $\rho(1, 1')$ is *N*-representable if and only if (iff), it corresponds to an antisymmetric *N*-body wavefunction $\Psi(\mathbf{1}\ldots\mathbf{N})$, which is a solution of the stationary Schrödinger equation:

$$\hat{H}\Psi(\mathbf{1}\ldots\mathbf{N}) = E\Psi(\mathbf{1}\ldots\mathbf{N}). \tag{1}$$

The one-body reduced density matrix is obtained from the wavefunction by integrating over all spin coordinates and all space coordinates but one:

$$\rho_1(\mathbf{1},\mathbf{1}') = N \sum_{\text{spins}} \int \Psi^*(\mathbf{1}\ldots\mathbf{N}) \Psi(\mathbf{1}'\ldots\mathbf{N}') d\mathbf{2}\ldots d\mathbf{N} \tag{2}$$

The spinless density matrix expanded in an orthonormal atomic orbital basis,

$$\rho_1(\mathbf{1},\mathbf{1'}) = \text{tr}\, \mathbf{P}\boldsymbol{\psi}_\Omega \boldsymbol{\psi}_\Omega^\dagger, \tag{3}$$

will correspond to an antisymmetric wavefunction $\Psi_{\text{det}}(\mathbf{1}\ldots\mathbf{N})$, where the subscript refers to a single determinantal expression of the wavefunction, iff,

$$\mathbf{P}^2 = \mathbf{P} \tag{4}$$

## 3.     The Clinton Equations

It is well known that the projector condition of Eq. (4) can be imposed upon the density matrix by obtaining $\mathbf{P}$ as a solution of the Clinton iterative matrix equations[19-22] of the general form

$$\mathbf{P}_{n+1} = 3\mathbf{P}_n^2 - 2\mathbf{P}_n^3 + \sum_k \lambda_k^n \mathbf{O}_k, \tag{5}$$

where $\mathbf{O}_k$ is the matrix representative of a quantum observable to be constrained by the solution $\mathbf{P}$ of the Clinton equations. Here the Lagrangian multipliers $\lambda$ are used to impose the satisfaction of X-ray constraints:

$$\text{tr}\, \mathbf{P}\mathbf{O}_k = \langle \hat{\mathbf{O}}_k \rangle \tag{6}$$

including normalization,

$$\text{tr}\, \mathbf{P} = N \tag{7}$$

and X-ray scattering factors,

$$\mathbf{F}(\mathbf{K}) = 2\,\text{tr}\, \mathbf{P}\mathbf{f}(\mathbf{k}) \tag{8}$$

where $\mathbf{f}(\mathbf{k})$ is a matrix of Fourier transforms of basis orbital products,

$$\mathbf{f}(\mathbf{K}) = \langle \boldsymbol{\psi}_\Omega | e^{i\mathbf{K}\cdot\mathbf{r}} | \boldsymbol{\psi}_\Omega \rangle. \tag{9}$$

For each scattering vector $\mathbf{K}$, if $\mathbf{P}^2 = \mathbf{P}$, then $\rho(1,1')$ as indicated in Eq. (3) is guaranteed to be single determinant $N$-representable.

## 4.    *N*-Representability: Small

For the case of fairly small molecular crystals, *i.e.*, on the order of 100 atoms or less, it is a straight forward expectation that the Clinton equations in the form of Eq. (5) can be applied using accurate diffraction data. Examples of application of the Clinton equations of X-ray data are well documented in the literature, see for example Ref. 23-28.

The result would be that an *N*-representable quantum mechanical density matrix may be extracted directly from the X-ray scattering data. That result is possible because of the important fact that for small molecules the number of linearly independent X-ray scattering data will greatly exceed the number of independent elements (unknowns) contained in the population matrix **P**.

There exist many early examples in which quantum mechanical projectors ($\mathbf{P}^2 = \mathbf{P}$) are extracted from X-ray scattering data.[23] As an illustration, the Clinton procedure applied to data from a berillium metallic crystal (P6$_3$/mmc) delivers excellent refinement statistics captured by an $R_f$ as low as 0.0018.[23] From such observations we may conclude that this approach is successful for small molecules.

## 5.    *N*-Representability: Big

As crystalline molecules under consideration grow in size, as will occur in systems of biological molecules, the number of X-ray data will be fewer that the number of the linearly independent elements of **P**, which are meant to be determined by that X-ray data.

This circumstance is inevitable as the growth in the number of elements in **P** increases as the square of the number of basis functions ($N^2$) while the growth in the X-ray data increases closer to the number of atoms in the asymmetric unit. At the cross over point in the growth of these numbers there is insufficient scattering data to determine the density matrix **P**. Obviously large molecules present a different case than do small molecules, and a different attack must be mounted to treat quantum mechanically the electronic structure of large molecules.

We have suggested[29,30] what is something like the opposite of the small molecule procedure. *Instead of extracting quantum mechanics from the scattering data, one may inject the quantum mechanics into the scattering experiment.* This requires breaking the X-ray experiment into two parts, which is made possible by invoking the Born-Oppenheimer approximation. We treat the two parts separately. The atomic structure is determined using a molecular description as a sum of multipole atomic densities for the large macromolecule. This gives an accurate set of atomic coordinates which defines the geometric structure. But the atomic coordinates are all that is needed to simply calculate, using quantum chemical formalism, the electronic structure of the molecular system.

In the Born-Oppenheimer approximation, the electronic wavefunction $\Psi_e$ satisfies the Schrödinger equation,

$$\hat{H}_e(\mathbf{r};\mathbf{R})\Psi_e(\mathbf{r};\mathbf{R}) = E_e \Psi_e(\mathbf{r};\mathbf{R}) \tag{10}$$

where $\mathbf{r}$ indicates the set of all position vectors of the electrons in the system, and $\mathbf{R}$ indicates all nuclear coordinates upon which the wavefunction depends parametrically. The nuclei vibrate each around its respective potential energy minimum formed by the electronic energy $E_e$. The electron density, which scatters the X-rays, is obtained from an integration over the square of the calculated wave function,

$$\rho(\mathbf{r}) = N \sum_{spin} \int \Psi^* \Psi \, d\mathbf{r}_2 d\mathbf{r}_3 \cdots d\mathbf{r}_N, \tag{11}$$

where integration is over all spins and all spatial coordinates but one.

The adequacy of the calculated electron density is adjudicated by construction of the crystallographic *R*-factor,

$$R = \frac{\sum |F_{obs.} - F_{calc.}|}{\sum |F_{obs.}|}, \tag{12}$$

which quantifies the agreement between the observed structure factors and those calculated from the model.

## 6.   The Kernel Energy Method (KEM)

The Born-Oppenheimer approximation has been invoked as a basis for breaking the X-ray problem into two parts, *viz.*, (1) the atomic structure is obtained using molecular multipole representations of atomic densities and (2) the electronic structure is obtained by the usual quantum chemical calculations. So now the question is what quantum method is convenient to use for the description of very large molecules? The Kernel Energy Method is here suggested.

The essential idea of this fragment method is to break a large molecule into mathematically tractable smaller pieces, call kernels. The total energy of the molecule is obtained as a KEM summation:[14,16,18,30-37]

$$E_{KEM} = \sum E_d - (n-2)\sum E_s,  \qquad (13)$$

where subscripts "d" and "s" refer to double and single kernels, respectively, and $n$ is the number of single kernels.

Many published calculations show that the KEM energies are quite accurate.[14,16,18,30-37] One example of such accuracy is the published calculation of the energy of a graphene flake under an external field reconstructed from KEM fragments[35] as illustrated in Figs. 1 and 2. Fig. 1 shows the graphene flake broken into three double and three single kernels and the recombination of their dipole moments *via* a summation similar to Eq. (13). Fig. 2 demonstrates the accuracy of KEM, not only for unperturbed ground state properties, but even in recovering response field-induced properties such as the change in the total energy and the induced dipole in the external field. Notice that as the imposed electric field increases the KEM energy tracks closely the "exact" energy trajectory.

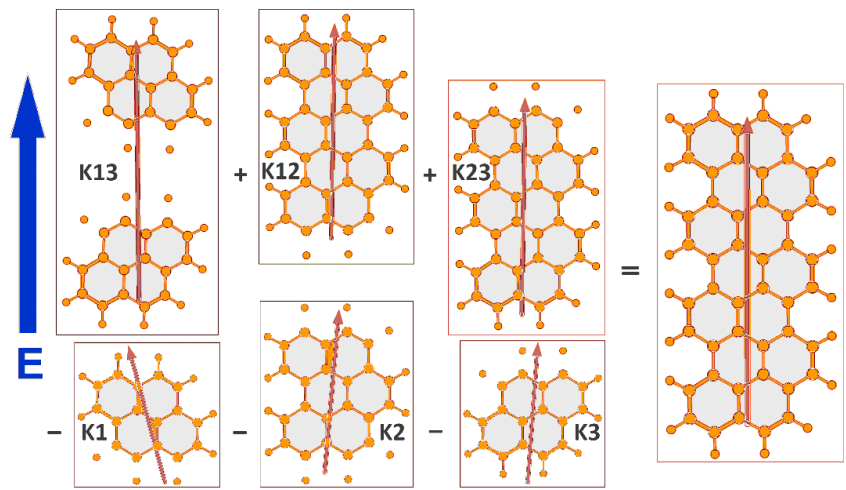

**Fig. 1**  The concept of KEM reconstruction illustrated using a vector property, that is, the induced dipole moment (the arrow on each kernel fragment) in an externally imposed electric field (the dark arrow on the left). Top row illustrates the vector sum of the induced dipole moment in the double kernels while the lower row subtracts the induced dipole moments in the single kernels (to correct for double counting). On the far right is an illustration of the large molecule target (a graphene nanoribbon with its overall induced dipole moment) according to the physicist's convention, *i.e.*, the head of the arrow points toward the positive side of the dipole.[38,39] (See Fig. 2 for the quantitative comparison of fragment reconstruction of total dipoles with the directly calculated dipoles under different external field strengths).

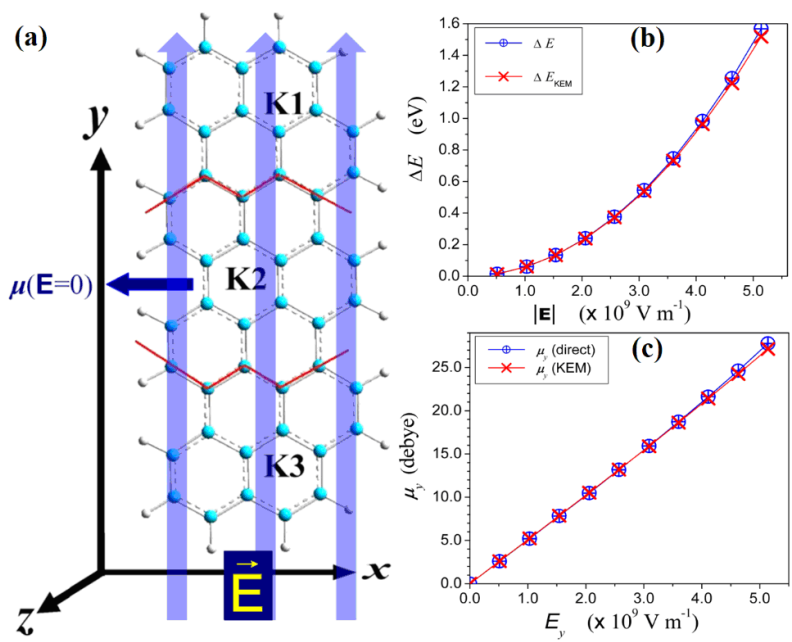

**Fig. 2**  (**a**) A ball-and-stick model of the graphene nanoribbon under an external electric field **E** (the long arrows along the *y*-axis). The zigzag (red) lines are those delimiting the kernel

fragments. The carbon atoms flanking the zigzag fission lines in the "environment/neighborhood" of a given kernel are replaced by hydrogen atoms (see Fig. 1 where the fission along the red line is implemented graphically). The small horizontal arrow is the intrinsic dipole of the ribbon in absence of any imposed fields pointing according to the physicist's convention.[38,39] (**b**) The energy of the full molecule at various magnitudes of |**E**| from a calculation on the full intact molecule and the one constructed from the fragments (according to Eq. (13)). (See Fig. 1 for the definition of the single and double kernel fragments). (**c**) A plot comparing the total induced molecular dipole reconstructed from KEM fragments (depicted in Fig. 1) and from a vector equation similar in form to Eq. (13) with that calculated directly.[35] (See Fig. 1).

Fig. 3 illustrates the fact that the KEM approximation does indeed apply to large biological macromolecules. In that case, a molecule consisting of more than 33,000 atoms has been calculated at the Hartree-Fock/STO3G level of theory.[33] We presume the method can be applied to ever larger and more complicated systems with millions of atoms.

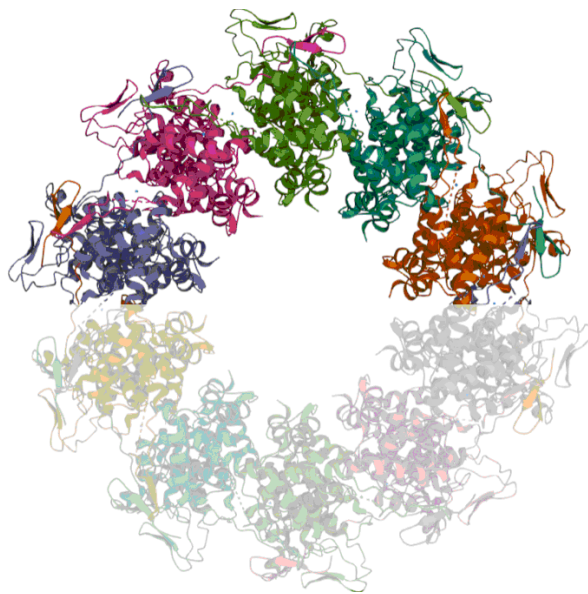

**Fig. 3** A ribbon representation of vesicular stomatitis virus nucleocapsid protein complex 10-mer (PDB code: 2QVJ). The asymmetric unit consists of five independent but strongly hydrogen-bonded peptides (33,175 atoms in total), the two asymmetric units composing the full protein are displayed with different shadings. The total energy of the asymmetric unit has been calculated at the Hartree-Fock/STO-3G level of theory[40] using the KEM approximation (Eq. (13)).[33] Corrections for the hydrogen-bonded interactions were added to the total energy using a correlated approach (MP2) and with a larger polarized basis set (6-31G(*d*,*p*)).[41] A comparison with the direct full-molecule calculation was not possible in this case due to its large size, but the point is to show that a heavily vetted methodology (namely KEM) is applicable to realistic large-sized bio-macromolecules.[33]

## 7. Formal and Real Computational scaling of the Kernel Energy Method

*Ab initio* calculations scale quickly as a power (generally greater than approximately three) of the number of basis functions associated with the chemical model. A simple algebraic manipulation for an idealized system can get us an order of magnitude (rough) estimate of the relative gains in CPU time achieved by the KEM partitioning. Assume that the size of the basis set required to describe the full large system is $M$ basis functions. Further, assume that the system is broken into $m$ kernel fragments each having the same number of basis functions $\mu$, and that the chosen chemical model chemistry scales as $\alpha$. With these simplifying idealized assumptions, the time of a KEM calculation relative to the fully-fledged calculation of the intact system ($t_{\text{rel}}$) is given by the ratio:

$$t_{\text{rel}} = \frac{t_{\text{KEM}}}{t_{\text{direct}}} = \frac{m\mu^\alpha + \frac{m^2 - m}{2}(2\mu)^\alpha}{M^\alpha}, \tag{14}$$

where the first term in the numerator is the total time (in arbitrary units) necessary to run calculations on $m$ equally-sized single kernels each described by $\mu$ basis functions and the second term in the numerator is the time necessary to run the calculations on $\frac{m^2 - m}{2}$ double kernels each having twice the number of basis functions as a single kernel. Since $M = m\mu$, we have:

$$t_{\text{rel}} = \frac{m\mu^\alpha + (m^2 - m)(2^{\alpha-1}\mu^\alpha)}{(m\mu)^\alpha}, \tag{15}$$

$$t_{\text{rel}} = \frac{2^{\alpha-1}(m-1) + 1}{m^{\alpha-1}}. \tag{16}$$

The result in this case is independent of the size of the basis set (which cancels in the derivation due to the identity of the number of basis functions in every single kernel). True this is an idealized state of affairs, but it does give the reader a good idea of the order of magnitude of the time saving advantage engendered by KEM.

Eq. (16) demonstrates that in order to achieve time saving, one must have at least three kernels, since splitting a system into only two kernels is meaningless as the double kernel itself is the full system. Hence, in this case, the KEM calculation will be even more costly (as predicted by the equation). Thus, there are thresholds of applicability of KEM but these thresholds are way lower than any imagined actual application of the method. The thresholds, for the idealized system described above, are: $m = 3$ kernels and $\alpha = 3$, values for which $t_{rel} = 1$ (*i.e.* the saving in time is zero). However, the ratio gets small (time saving gets large) very fast as can be seen from Table 1 which lists a few values of the relative time within the idealized system that leads to Eq. (16). From the Table, we can see that, with a cubic scaling computational method and when the system is broken to $\approx 400$ single kernels, the computational time is cut by two orders of magnitude. Now if the scaling is to the 4$^{th}$ power, that saving is now between 4 and 5 orders of magnitude. The saving increases fast with the scaling of the computational method as seen, for example for $\alpha = 5$, where now a 400 kernel fragmentation is 10 million times faster that the full molecular calculation. Fig. 4 displays the relative times as a function of the number of kernels for three exponential computational scale factors, $\alpha$, showing the fast decline of $t_{rel}$.

To illustrate the time saving effected by the KEM approach, suppose we have a peptide made up 100 amino acids, whereby an average amino acid has around 20 atoms (including hydrogen atoms). A moderate basis set would probably average to say 10 basis functions per atom, that is, the entire peptide would require of the order of $100 \times 20 \times 10 = 20,000$ basis functions. Suppose we use a single determinant method to perform an *ab initio* calculation on this system and that the calculation scales, say, cubically. In this case, the computational effort, in absence of parallelization, on a given CPU would be of the order of $(20,000)^3 = 8 \times 10^{12}$ time units.

Now imagine breaking this system according to the KEM (Eq. (13)) into 100 kernels each consisting of a single amino acid (with 20 atoms each) and all possible unique 4950 (= $(100^2-100)/2$) double kernels of 40 atoms each. Using the same CPU and computational conditions/environment this would lead to approximately $100 \times 200^3 + 5000 \times 400^3 \approx 3 \times 10^{11}$ time units, which is about 4% of the CPU time necessary for the full molecule, as can be seen from the row of 96 ($\approx 100$) kernels and column $\alpha = 3$ of Table 1.

The saving, of course, is quickly magnified with the power dependence of the computational difficulty. A glance at Eq. (16) shows the reason for this power-dependence and that is that the numerator (the computational demand on KEM) scales linearly with $m$ and only the number "2" scales as $(\alpha - 1)$, while the denominator scales as $m^{\alpha-1}$ with $m \gg 2$. For example, in the illustrative example outlined above, if now the calculation scales to the power of 4, the saving without parallelization on the same system becomes ca. $10^{17}/10^{14} \approx 0.08\%$. Clearly, KEM can dramatically speed-up calculations especially for large systems and for more sophisticated calculations that scale at higher exponents.

Now suppose we can parallelize this on 100 CPU (a modest computational cluster by today's standards), this improves time savings by two further orders of magnitude.

**Table 1.** Relative time as a function of both the number of single kernels and the computational model chemistry scaling parameter ($t_{rel}(m,\alpha)$) for a few representative values of $m, \alpha$ (according to the idealized model (Eq. 16).

| $m$ | $t_{rel}(m,\alpha)$ | | |
|---|---|---|---|
| | $\alpha = 3$ | $\alpha = 4$ | $\alpha = 5$ |
| 3 | 1 | $6\times10^{-1}$ | $4\times10^{-1}$ |
| 6 | $6\times10^{-1}$ | $2\times10^{-1}$ | $6\times10^{-2}$ |
| 12 | $3\times10^{-1}$ | $5\times10^{-2}$ | $9\times10^{-3}$ |
| 24 | $2\times10^{-1}$ | $1\times10^{-2}$ | $1\times10^{-3}$ |
| 48 | $8\times10^{-2}$ | $3\times10^{-3}$ | $1\times10^{-4}$ |
| 96 | $4\times10^{-2}$ | $9\times10^{-4}$ | $2\times10^{-5}$ |
| 192 | $2\times10^{-2}$ | $2\times10^{-4}$ | $2\times10^{-6}$ |
| 384 | $1\times10^{-2}$ | $5\times10^{-5}$ | $3\times10^{-7}$ |

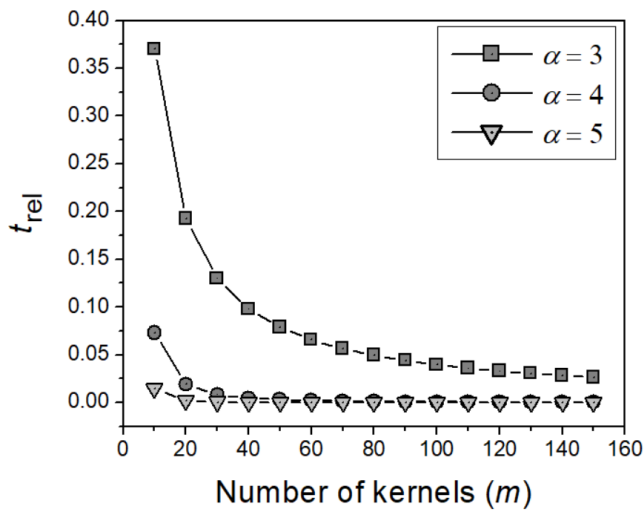

**Fig. 4** Theoretical relative time (ratio of time of a KEM calculation with respect to the full molecule (direct) calculation) according to Eq. (16) for three scaling factors as a function of the number of single kernels into which the large molecule is broken.

## 8. *N*-Representable Kernel Energy Method

But a good energy, while desirable, is not entirely what is required of a quantum mechanical result. It is also required that the electron density and the corresponding density matrices must be *N*-representable. It is this mathematical requirement that guarantees correspondence to the experimental indistinguishability of the electrons.

It occurs that there is no mandate for KEM results to be *N*-representable. However, a recent publication[29] shows a procedure by which *N*-representability can be imposed upon the results of KEM calculations.

As usual the Clinton equations are used to bring together KEM density matrices and single Slater determinants from which they are derived. This occurs as follows. The quantum kernels of a molecule are calculated in the usual way. Each kernel K will have density matrix **R** corresponding to an atomic orbital basis $\psi$, *i.e.*,

$$\rho_{1(K)}(\mathbf{r},\mathbf{r}') = 2\, tr\, \mathbf{R}\, \psi(\mathbf{r})\psi^{\dagger}(\mathbf{r}') . \tag{17}$$

The collection of density matrices **R**, for all double and single kernels are collected together to form one large density matrix $\mathbf{R}_{KEM}$ representing the full molecule, thus

$$\mathbf{R}_{KEM} \equiv \sum \mathbf{R}_d^{aug} - (n-2)\sum \mathbf{R}_s^{aug} , \tag{18}$$

where the density matrices for double and single kernels are indicated in an obvious use of subscripts.

Each double and single kernel density matrix would be of dimensions much smaller than the density matrix of the full molecule. The superscript "aug" indicates the dimensions of each kernel density matrix is built-up to the dimension of the full molecule density matrix by augmenting the kernels with a sufficient number of zero elements. In that way, every matrix in Eq. (18) is of the same dimension, *viz*., that of the full molecule. $\mathbf{R}_{KEM}$ is therefore the density matrix of the full molecule in KEM form.

The matrix $\mathbf{R}_{KEM}$ corresponds to a basis of atomic orbitals $\psi$, the collection of all MO's that were used to carry out the original kernels calculations. The full molecule density matrix in that basis, is

$$\rho_{1(KEM)}(\mathbf{r},\mathbf{r'}) = 2tr\,\mathbf{R}_{KEM}\,\psi(\mathbf{r})\psi^{\dagger}(\mathbf{r'}) . \tag{19}$$

Using the AO overlap matrix

$$\mathbf{S} \equiv \int \psi(\mathbf{r})\cdot\psi^{\dagger}(\mathbf{r'})d\mathbf{r} . \tag{20}$$

The basis is transformed to an orthonormal atomic basis according to the Löwdin procedure:[42]

$$\psi_{\Omega} = \mathbf{S}^{-1/2}\psi . \tag{21}$$

The density matrix is there by transformed into

$$\mathbf{P}_{0} \equiv \mathbf{S}^{1/2}\mathbf{R}_{KEM}\mathbf{S}^{1/2} . \tag{22}$$

The subscript zero is used to indicate that this matrix will by used as the initial iterant in the Clinton equations (Eqs. (5 - 7)). The result is a projector $\mathbf{P}$ which represents the density matrix of the full molecule in the orthonormal basis, *i.e.*,

$$\rho_{1(KEM)}(\mathbf{r},\mathbf{r'}) = 2tr\,\mathbf{P}\,\psi_{\Omega}(\mathbf{r})\psi_{\Omega}^{\dagger}(\mathbf{r'}) . \tag{23}$$

This density matrix is *N*-representable because $\mathbf{P}$ is a normalized projector.

The value of imposing *N*-representability upon the KEM density matrix may be noticed by results obtained with a series of water molecules. Gadre *et al.* energy optimized collections of water molecules of size 3 up to 20.[43] These collections of optimized water molecules are used as test cases to ascertain the importance of imposed *N*-representability upon KEM results.

In Table 2, numerical results extracted from the recent literature[29] are shown for water molecules taken in group of 10, 12, 14, 15, 16 and 20. For each of these cases are shown the differences $E_{KEM} - E_{full}$ and $E_{[\mathbf{P}_{projector}]} - E_{full}$. The column showing the energy differences without imposing *N*-representability upon the KEM energy $E_{KEM}$ indicates four cases for which $E_{KEM} - E_{full}$ is a negative number. Such negative differences correspond to a violation of the variational principle. This is a certain indication that *N*-representability is violated. The two entries at the bottom of that same column show relatively large differences (errors) in the KEM calculations. In the last column, the differences correspond to enforcing *N*-representability onto the KEM calculations. All entries in that column are positive, as must be the case, with *N*-representable density matrices. Notice also all the *N*-representability difference errors are generally smaller than those in the previously discussed column (as can be gauged by the average absolute errors at the bottom of the Table: 2.8±2.5 and 1.2±0.5 kcal/mol for the non-*N*-representable and the *N*-representable calculations, respectively). Hence, as expected, *N*-representability enforces the variational theorem and, moreover, tends to reduce KEM errors.[29]

**Table 2.** Error in the KEM total energies with and without the imposition of *N*-representability. (Based on results reported in Ref. 29 at the RHF/6-31G level of theory).

| $n_{water}$ | $E_{KEM} - E_{full}$ (kcal/mol) | $E_{[\mathbf{P}_{projector}]} - E_{full}$ (kcal/mol) |
|---|---|---|
| 10 | -0.4 | 0.7 |
| 12 | -3.8 | 0.8 |
| 14 | -0.5 | 0.8 |
| 15 | -1.0 | 1.5 |
| 16 | 6.4 | 1.9 |
| 20 | 4.4 | 1.6 |
| A.a.e.* | 2.8 ± 2.5 | 1.2 ± 0.5 |

* A.a.e. stands for the average absolute error (± standard deviation).

## 9. Breaking P into Kernels P′

The kernel energy method introduces quantum mechanics into a large system of known geometry obtained, for example, from crystallography. KEM then fleshes it out to deliver quantum mechanical density and density matrices. The procedure, as already shown above, can affect huge savings in CPU times. By constraining the electron density and density matrices to be *N*-representable one can *reverse* the procedure whereby KEM can be used in conjunction with the crystallographic refinement to deliver quantum mechanical properties such as momentum densities or energies.[17,44]

Now we discuss the second task (reversing the procedure) alluded-to above. Beginning with a projector **P**, that carries the *N*-representability of the full system, break that matrix into a KEM summation of kennel subspecies **P′** preserving *N*-representability. That is to say, find the kernel matrices **P′** that satisfy,

$$\mathbf{P} = \sum \mathbf{P}'_d - (n-2) \sum \mathbf{P}'_s . \tag{24}$$

Each of the projector matrices **P′** on the right side of the last equation belong to a subspace of the projector **P** representing the full system. Eq. (24) is the transformation of Eq. (18) into an orthonormal atomic basis, as in Eq. (21). The kernels **P′** which satisfy Eq. (24) will preserve the *N*-representability carried by the known projector **P**.

To find the set of kernel projectors **P′** invokes the triple product **P′PP′**. In general this triple product of projectors is not itself a projector. But it is instructive to recognize that it is an expansion onto the subspace defined by **P′**. For example, consider matrices defined as,

$$\mathbf{P} = \begin{pmatrix} x & x & x \\ x & x & x \\ x & x & x \end{pmatrix}, \tag{25}$$

and

$$\mathbf{P}' = \begin{pmatrix} y & y & 0 \\ y & y & 0 \\ 0 & 0 & 0 \end{pmatrix} . \tag{26}$$

Notice that the triple product of projectors reduces to the space of **P′**, *i.e.*, matrix multiplication yields

$$\mathbf{P'PP'} = \begin{pmatrix} z & z & 0 \\ z & z & 0 \\ 0 & 0 & 0 \end{pmatrix} \quad (27)$$

The generalized matrix elements *x, y, z*, may be matrices themselves. They are used here *simply and only* as *place holders* indicating *only* the size of the spaces onto which the matrices act. The important thing to notice is that the space of triple product **P′PP′** reduces that of **P** to that of its subspace **P′**, as dictated by the rules of matrix multiplication.

In order for **P′** to be a subspace belonging entirely to **P**, we must have:

$$\mathbf{P'} = \mathbf{P'PP'}. \quad (28)$$

The Clinton equations allow for a search of that set of matrices **P′** which satisfy Eq. (28). The constraints applied in the Clinton equations for determination of each kernel subspace **P′** would be first, normalization of **P′**, and second, satisfaction of Eq. (28).

## 10. Details for Determination of Constraint **P′** = **P′PP′**

A necessary and sufficient condition for the satisfaction of this constraint is,

$$\mathrm{tr}\,(\mathbf{P'} - \mathbf{P'PP'})^2 = 0. \quad (29)$$

Expanding the square, and solving for $\mathrm{tr}\,(\mathbf{P'})^2$, we get:

$$\mathrm{tr}\,(\mathbf{P'P'}) = \mathrm{tr}[(\mathbf{P'})^2\mathbf{PP'} + \mathbf{P'P}(\mathbf{P'})^2 - (\mathbf{P'PP'})^2] \quad (30)$$

The standard form of a Clinton equation constraint is:

$$\mathrm{tr}\,(\mathbf{PO}) = \langle \hat{O} \rangle \quad (31)$$

thus we define

$$\mathbf{O} \equiv \mathbf{P'} \quad (32)$$

and,

$$\langle \hat{O} \rangle = \mathrm{tr}[(\mathbf{P'})^2\mathbf{PP'} + \mathbf{P'P}(\mathbf{P'})^2 - (\mathbf{P'PP'})^2] \quad (33)$$

This result represents the constraint needed to apply the Clinton equations for determination of **P′**.

Applying the Clinton equations using the two constraints of normalization Eq. (7) and **P′** = **P′PP′** Eqs. (29 - 33) delivers the kernel subspaces of the full system's projector **P**, that is, Eq. (24).

## 11. Conclusions

The Clinton equations deliver X-ray $N$-representability of the experimentally-consistent density matrices for systems *big* and *small*. For a small system, we *extract* the quantum mechanics from X-ray data. In contrast, for a large system, that is, a system with less data than necessary for a statistically-sound representation of the independent parameters defining the density matrix, we *inject* quantum mechanics into X-ray data. The philosophies proceed in opposite directions but they are complementary. In either case, *the Clinton equations are "means" to the* N-*representability "end"*.

**Acknowledgments**

The authors are grateful to Dr. Walter Polkosnik for valuable conversations. Funding for this project was provided by the *U.S. Naval Research Laboratory* (project # 47203-00 01) and by a *PSC CUNY Award* (project # 63842-00 41), the *Natural Sciences and Engineering Research Council of Canada* (NSERC), *Canada Foundation for Innovation* (CFI), and *Mount Saint Vincent University*.

**References**

1. L. Pauling, E. B. Jr. Wilson, Introduction to Quantum Mechanics with Applications to Chemistry; Dover Publications, Inc.: New York, **1963**.

2. L. Pauling. The nature of the chemical bond. Application of the results obtained from the quantum mechanics and from a theory of paramagnetic susceptibility to the structure of molecules. *J Am. Chem. Soc.* **1931**, *53*, 1367-1400.

3. S. Shaik, P. C. Hiberty, A Chemist's Guide to Valence Bond Theory; John Wiley and Sons, Inc.: New Jersey, **2007**.

4. L. Pauling, The Nature of the Chemical Bond (Third Ed.); Cornell University Press: Ithaca, N.Y., **1960**.

5. A. Rich, N. Davidson, Structural Chemistry and Molecular Biology (A Volume dedicated to Linus Pauling by his Students, Colleagues, and friends); W. H. Freeman and Company: San Francisco and London, **1968**.

6. A. Genoni, L. Bucinsky, N. Claiser, J. Contreras-García, B. Dittrich, P. M. Dominiak, E. Espinosa, C. Gatti, P. Giannozzi, J.-M. Gillet, D. Jayatilaka, P. Macchi, A. Ø. Madsen, L. Massa, C. F. Matta, K. M. Merz Jr., P. Nakashima, H. Ott, U. Ryde, W. Scherer, K. Schwarz, M. Sierka, S. Grabowsky. Quantum crystallography: Current developments and future perspectives. *Chem. Eur. J.* **2018**, *24*, 10881-10905.

7. S. Grabowsky, A. Genoni, H.-B. Bürgi. Quantum crystallography. *Chem. Sci.* **2017**, *8*, 4159-4176.

8. P. Macchi. The connubium between crystallography and quantum mechanics. *Crystallogr. Rev.* **2020**, *26*, 209-268.

9. C. Gatti, P. Macchi, Modern Charge-Density Analysis; Springer: Berlin, **2012**.

10. V. Tsirelson. Early days of quantum crystallography: A personal account. *J. Comput. Chem.* **2018**, *39*, 1029-1037.

11. D. Jayatilaka. Wave function for beryllium from X-ray diffraction data. *Phys. Rev. Lett.* **1998**, *80*, 798-801.

12. D. Jayatilaka, D. J. Grimwood. Wavefunctions derived from experiment. I. Motivation and theory. *Acta Cryst. A* **2001**, *57*, 76-86.

13. D. Jayatilaka, D. J. Grimwood. Wavefunctions derived from experiment. II. A wavefunction for oxalic acid dihydrate. *Acta Cryst. A* **2001**, *57*, 87-100.

14. Huang L.; Massa, L.; Karle, J. Chapter 1 in: *Quantum Biochemistry: Electronic Structure and Biological Activity (Vol. I)*; Matta C. F. (Ed.), Wiley-VCH: Weinheim, 2010; pp. 3-60.

15. L. Huang, L. Massa, J. Karle. Quantum crystallography. *J. Mol. Struct.* **1999**, *474*, 9-12.

16. L. Massa, L. Huang, J. Karle. Quantum crystallography and the use of kernel projector matrices. *Int. J. Quantum. Chem.* **1995**, *56*, 371-384.

17. L. Massa, C. F. Matta. Exploiting the full quantum crystallography. *Can. J. Chem.* **2018**, *96*, 599-605.

18. L. Massa, C. F. Matta. Quantum cystallography: A perspective. *J. Comput. Chem.* **2018**, *39*, 1021-1028.

19. W. L. Clinton, C. A. Frishberg, L. J. Massa, P. A. Oldfield. Methods for obtaining an electron-density matrix from x-ray data. *Int. J. Quantum Chem.* **1973**, *7*, 505-514.

20. L. J. Massa, W. L. Clinton. Antisymmetric wavefunction densities from coherent diffraction data. *Trans. Am. Cryst. Assoc.* **1972**, *8*, 149-153.

21. W. L. Clinton, L. J. Massa. Determination of the electron density matrix from X-ray diffraction data. *Phys. Rev. Lett.* **1972**, *29*, 1363-1366.

22. W. L. Clinton, A. J. Galli, L. J. Massa. Direct determination of pure-state density matrices. II. Construction of constrained idempotent one-body densities. *Phys. Rev.* **1969**, *177*, 7-12.

23. L. Massa, M. Goldberg, C. Frishberg, R. Boehme, S. LaPlaca. Wave functions derived by quantum modeling of the electron density from coherent X-Ray diffraction: Beryllium metal. *Phys. Rev. Lett.* **1985**, *55*, 622-625.

24. R. F. Boehme, S. J. La Placa. Empirical molecular hydrogen wave function modeled from theoretically derived x-ray diffraction data. *Phys. Rev. Lett.* **1987**, *59*, 985-987.

25. Y. V. Aleksandrov, V. G. Tsirelson, I. M. Reznik, R. P. Ozerov. The Crystal Electron Energy and Compton Profile Calculations from X-Ray Diffraction Data. *Phys. Stat. Sol. (b)* **1989**, *155*, 201-207.

26. S. T. Howard, J. P. Huke, P. R. Mallinson, C. S. Frampton. Density-matrix refinement for molecular crystals. *Phys. Rev. B* **1994**, *49*, 7124-7136.

27. J. A. Snyder, E. D. Stevens. A wavefunction and energy of the azide ion in potassium azide obtained by a quantum-mechanically constrained fit to X-ray diffraction data. *Chem. Phys. Lett.* **1999**, *313*, 293-298.

28. L. Huang, L. Massa, J. Karle. Quantum crystallography applied to crystalline maleic anhydride. *Int. J. Quantum Chem.* **1999**, *73*, 439-450.

29. W. Polkosnik, L. Massa. Single determinant *N*-representability and the kernel energy method applied to water clusters. *J. Comput. Chem.* **2018**, *39*, 1038-1043.

30. W. Polkosnik, C. F. Matta, L. Huang, L. Massa. Fast quantum crystallography. *Int. J. Quantum Chem.* **2019**, *119*, Article # e26095 (pp. 1-11).

31. L. Huang, L. Massa, J. Karle. Kernel projector matrices for Leu[1]-zervamicin. *Int. J. Quantum. Chem.* **1996**, *60*, 1691-1700.

32. L. Huang, L. Massa, J. Karle. Kernel energy method: Application to insulin. *Proc. Natl. Acad. Sci. USA* **2005**, *102*, 12690-12693.

33. L. Huang, L. Massa, J. Karle. Kernel energy method applied to vesicular stomatitis virus nucleoprotein. *Proc. Natl. Acad. Sci. USA* **2009**, *106*, 1731-1736.

34. L. Huang, H. Bohorquez, C. F. Matta, L. Massa. The kernel energy method: Application to graphene and extended aromatics. *Int. J. Quantum Chem.* **2011**, *111*, 4150-4157.

35. L. Huang, L. Massa, C. F. Matta. A graphene flake under external electric fields reconstructed from field-perturbed kernels. *Carbon* **2014**, *76*, 310-320.

36. M. J. Timm, C. F. Matta, L. Massa, L. Huang. The localization-delocalization matrix and the electron density-weighted connectivity matrix of a finite graphene nanoribbon reconstructed from kernel fragments. *J. Phys. Chem. A* **2014**, *118*, 11304-11316.

37. L. Huang, C. F. Matta, L. Massa. The kernel energy method (KEM) delivers fast and accurate QTAIM electrostatic charge for atoms in large molecules. *Struct. Chem.* **2015**, *26*, 1433-1442.

38. C. A. Coulson, Electricity; Oliver and Boyd: London, **1961**.
39. J. W. Hovick, J. C. Poler. Misconceptions in sign conventions: Flipping the electric dipole moment. *J. Chem. Edu.* **2005**, *82*, 889.
40. T. Clark, A Handbook of Computational Chemistry; John Wiley and Sons: New York, **1985**.
41. A. Szabo, N. S. Ostlund, Modern Quantum Chemistry: Introduction to Advanced Electronic Structure Theory; Dover Publications, Inc.: New York, **1989**.
42. P.-O. Löwdin. On the Nonorthogonality Problem. *Adv. Quantum Chem.* **1970**, *5*, 185-199.
43. S. Maheshwary, N. Patel, N. Sathyamurthy, A. D. Kulkarni, S. R. Gadre. Structure and stability of water clusters $(H_2O)_n$, $n$ = 8 - 20: An *ab initio* investigation. *J. Phys. Chem. A* **2001**, *105*, 10525-10537.
44. L. Massa, T. Keith, Y. Cheng, C. F. Matta. The kernel energy method applied to quantum theory of atoms in molecules – energies of interacting quantum atoms. *Chem. Phys. Lett.* **2019**, *734*, Article #136650 (pp. 1-4).